\documentclass{jpconf}
\usepackage{graphicx,amssymb}

\begin{document}
\title{Searching for stochastic gravitational-wave background 
with the co-located LIGO interferometers}

\author{Nickolas V Fotopoulos
(for the LIGO Scientific Collaboration)}

\address{University of Wisconsin-Milwaukee, 1900 E Kenwood Blvd Rm 442, Milwaukee, WI, USA}
\ead{nvf@gravity.phys.uwm.edu}

\begin{abstract}

This paper presents techniques developed by the LIGO Scientic Collaboration
to search for the stochastic gravitational-wave background using the
co-located pair of LIGO interferometers at Hanford, WA. We use correlations
between interferometers and environment monitoring instruments, as well as
time-shifts between two interferometers (described here for the first time)
to identify correlated noise from non-gravitational sources. We veto
particularly noisy frequency bands and assess the level of residual
non-gravitational coupling that exists in the surviving data.
\end{abstract}

\section{Introduction}

The LIGO (Laser Interferometer Gravitational-Wave Observatory) project has constructed two detectors,
dubbed H1 and H2, at its Hanford observatory (on the US Department of
Energy's Hanford Site in Washington, USA) and a third, L1, at its Livingston observatory
(in Livingston Parish, Louisiana, USA).  These detectors are Fabry-Perot, power-recycled Michelson
interferometers that detect minute changes in the differential lengths of their arms, the signatures of
incident gravitational waves. The Fabry-Perot cavities, or arms, are \mbox{4\,km} long in the H1 and L1
detectors and \mbox{2\,km} long in H2.  In November 2005, the LIGO instruments achieved their design
sensitivities~\cite{ScienceRuns} and began taking data for LIGO's fifth science run, S5.  The S5
observing effort continued through September 2007 and for some of its duration included the GEO 600
and VIRGO instruments. This run enables searches for different types 
of gravitational radiation with unprecedented sensitivities.

One of the LSC's (LIGO Scientific Collaboration's) major efforts is to search for an SGWB (stochastic gravitational-wave background),
which is the gravitational analogue to the electromagnetic CMB (cosmic microwave background).  The SGWB is
predicted to contain contributions from unresolved populations of distant astrophysical processes
involving compact objects as well as from early cosmological phenomena such as inflation and cosmic
string collisions.  The LSC has not yet finalized results for an SGWB search of the S5 data, but has published
upper limits using S4 and earlier data sets~\cite{S4paper}\cite{S3paper}\cite{S1paper}.

\subsection{The stochastic search and H1-H2}

The SGWB is usually described in terms of the GW spectrum:
\begin{equation}
    \Omega_\mathrm{GW}(f) \, \frac{\rmd f}{f} \equiv \frac{\rmd\rho_\mathrm{GW}}{\rho_\mathrm{c}} ,
\end{equation}
where $\rmd\rho_{\rm GW}$ is the energy density of gravitational radiation
contained in the frequency range $f$ to $f+\rmd f$ \cite{AllenRomano97},
$\rho_\mathrm{c}$ is the critical energy density of the Universe, and $f$ is
frequency.

\begin{figure}
\includegraphics[width=3.1in]{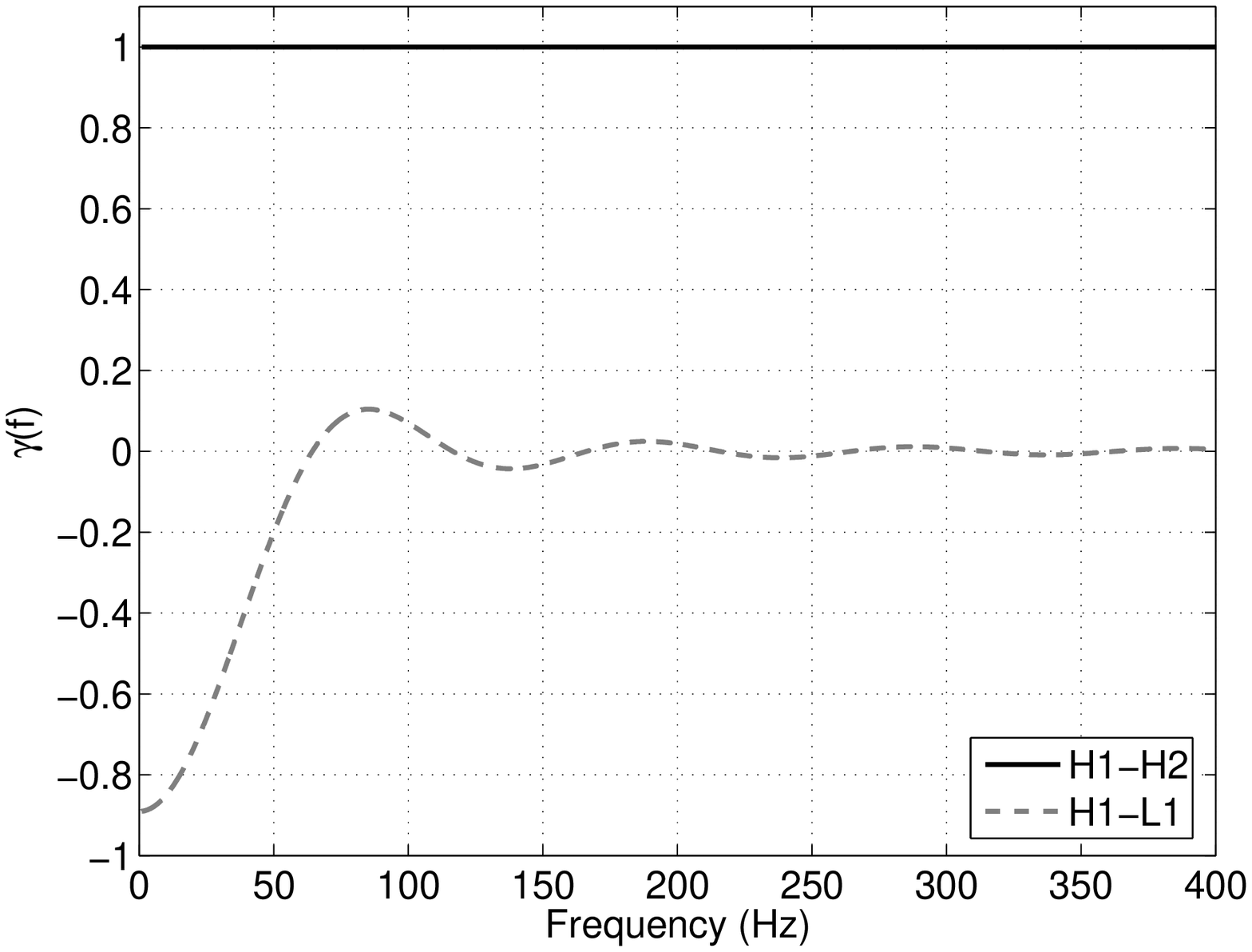}
\includegraphics[width=3.1in]{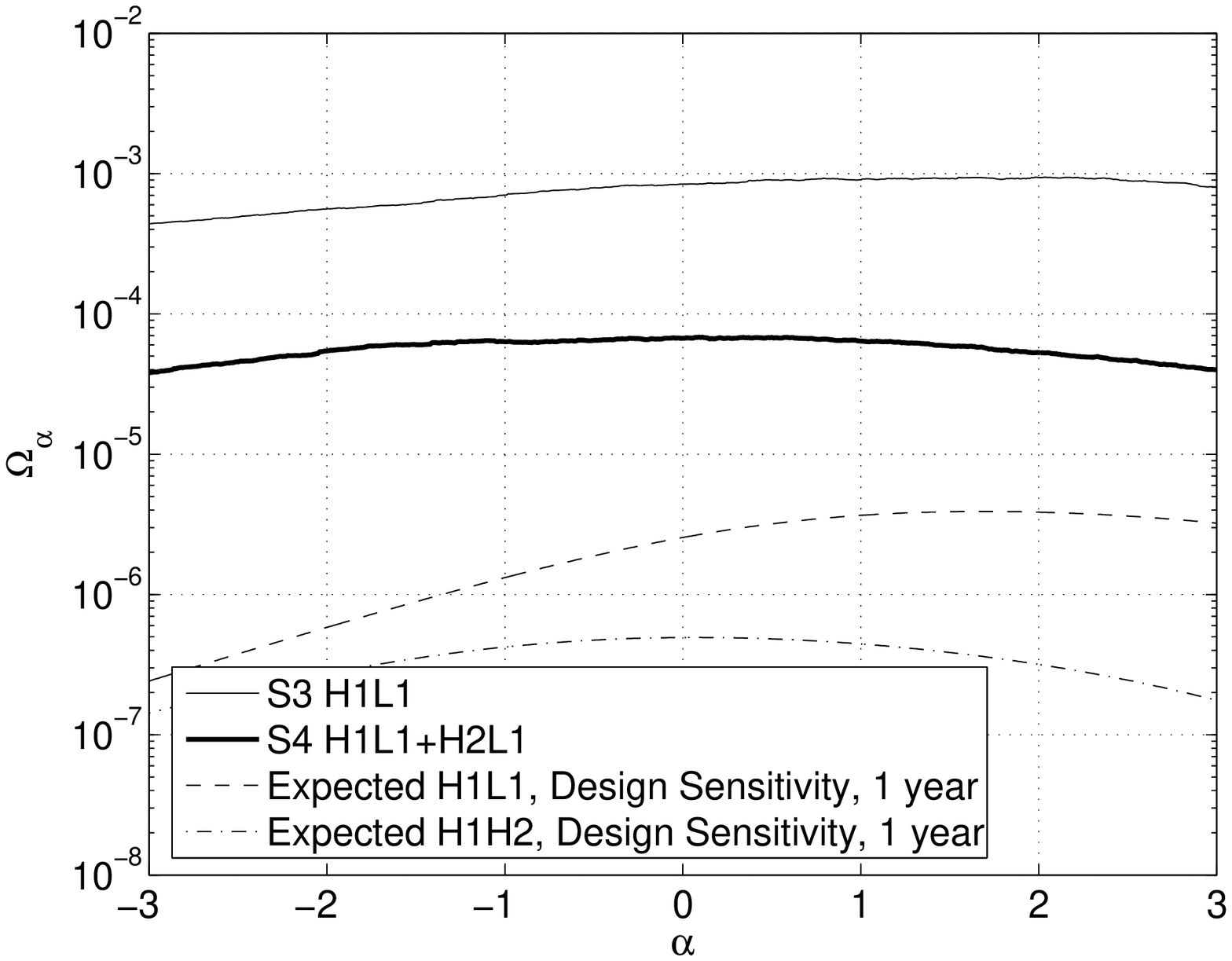}
\caption{Left: Overlap reduction for the H1-H2
(black solid) and for the H1-L1 pairs (gray dashed). Right: 
Most recent upper limits (S3 (thin solid) and S4 (thick solid) runs) 
on $\Omega_{\alpha}$ as 
a function of $\alpha$, and sensitivity predictions for
S5 for the H1-L1 (dashed) and H1-H2 (dot-dashed) pairs, 
assuming no instrumental correlations, design sensitivity, 
and one year of integration.}
\label{overlap}
\end{figure}

LSC searches for the isotropic component of the SGWB deploy the cross-correlation
technique described in Allen and Romano\cite{AllenRomano97}.
The technique estimates the contribution to the detector noise due to the SGWB
by cross-correlating the detector output between a pair of detectors;  if the detectors'
intrinsic noises are uncorrelated, the common astrophysical component can be distinguished.  We
define the cross-correlation estimator:
\begin{eqnarray}
Y = \int_{0}^{+\infty } \rmd f \; Y(f) =
\int_{-\infty }^{+\infty } \rmd f \int_{-\infty }^{+\infty } \rmd f' \;
\delta_T (f-f') 
\; \tilde{s}_1^{*}(f) \; \tilde{s}_2(f') \; \tilde{Q}(f')\,,
\label{ptest}
\end{eqnarray}
where $\delta_T$ is a finite-time approximation to the Dirac delta function,
$\tilde{s}_1$ and $\tilde{s}_2$ are the Fourier transforms of the
strain time-series of two interferometers, and $\tilde{Q}$ is
a filter function. In the limit when the detector noise is much larger 
than the GW signal, and assuming that the detector noise is stationary, 
Gaussian, and uncorrelated between the two interferometers, the
variance of the estimator $Y$ is given by:
\begin{eqnarray}
\sigma_Y^2 = \int_0^{+\infty} \rmd f \; \sigma_Y^2(f) \approx 
\frac{T}{2} \int_0^{+\infty} \rmd f P_1(f) P_2(f)
| \tilde{Q}(f) |^2\,,
\label{sigma}
\end{eqnarray}
where the $P_i(f)$ are the one-sided power spectral densities (PSDs) 
of the two interferometers
and $T$ is the (single-segment) integration time. Optimization of the signal-to-noise
ratio leads to the following form of the 
optimal filter \cite{AllenRomano97}:
\begin{eqnarray}
\tilde{Q}(f) = \mathcal{N} \; \frac{\gamma(f) 
S_{\rm GW}(f)}{P_1(f) P_2(f)} {\rm \; , \;\;\;\; where \;\;\;}
S_{\rm GW}(f) = \frac{3 H_0^2}{10 \pi^2} \; 
\frac{\Omega_{\rm GW}(f)}{f^3} \; .
\label{optfilt}
\end{eqnarray}
Here, $S_{\rm GW}(f)$ is 
the strain power spectrum of the SGWB for which we are searching. 
Assuming a power-law template GW spectrum with index $\alpha$,
$\Omega_{\rm GW}(f) = \Omega_{\alpha} (f / 100 {\rm \, Hz})^{\alpha}$,
the normalization constant $\mathcal{N}$
is chosen such that $\langle Y\rangle = \Omega_{\alpha} T$.
Finally, $\gamma(f)$ is the overlap reduction function. It captures the signal
reduction due to two effects: (i) relative translation between interferometers
introduces a frequency- and direction-dependent phase difference, and (ii)
relative rotation makes each detector sensitive to a different polarization 
of incident radiation. As shown
in figure~\ref{overlap}, the identical antenna patterns of the 
collocated Hanford interferometers imply $\gamma(f) = 1$, while for 
the H1-L1 pair, the overlap reduction is significantly lower than 1 above \mbox{50\,Hz}. 
In the most common search case, which assumes a frequency-independent spectrum ($\alpha=0$), 
the theoretical advantage for H1-H2 over H1-L1 is approximately a factor
of ten in $\Omega_0$, as shown in figure~\ref{overlap}.
This advantage could be eroded, however, by instrumental correlations between H1 and H2.

Since H1 and H2 share the same vacuum enclosures and many of their 
optical benches share the same building, they are susceptible to common 
environmental disturbances.  The
cross-correlation induced by the environment could be greater than the
noise floor of our
measurements, and could mimic, mask, or otherwise cause us to misestimate the amplitude or shape of a gravitational-wave signal.  
Hence, the H1-H2 pair has not been used for the SGWB searches so far.

We have developed two techniques to identify non-gravitational-wave 
correlations between
H1 and H2 data.  The IFO-PEM coherence
technique was discussed in~\cite{IFOPEMcoherence}. Here,
IFO is shorthand for ``interferometer'' and PEM is an acronym
for ``physical environmental monitor''.  There are numerous sensors situated throughout the LIGO
observatories providing data to the PEM channels; there are accelerometers, magnetometers, seismometers, voltmeters, and anemometers.
By computing the coherence spectrum between each IFO channel and each PEM channel, one can find
the linear environmental coupling at each frequency.  The technique is sensitive to features of arbitrary bandwidths
and resolutions, but is restricted to linear couplings and obviously cannot give information where PEM
coverage is non-existent.

Section~\ref{sec:time_shift} of this paper will describe the second, complementary handle on
non-gravitational coupling, the time-shift technique.  The time-shift technique's added value is that
it is not hindered by incomplete knowledge of the environment; its trade-off is in its narrow bandwidth.
Section~\ref{sec:relating_techniques} will connect
the two methods to each other and to the measured quantities in the SGWB search and
section~\ref{sec:search_algorithm} will describe how we envision applying these techniques together to
extract the best possible SGWB measurement from the H1-H2 pair.  The paper will close with a statement of where we stand on the H1-H2 problem.

\section{Time-shift technique} \label{sec:time_shift}

The main idea of the time-shift technique is to perform the search 
described in the previous section after introducing an unphysical 
time-shift between the two interferometers' data-streams. That is, since the SGWB is expected
to be broadband (at least hundreds of Hz wide), the SGWB-induced cross-correlation
between the two interferometers is expected to disappear after a time-shift $\tau$
larger than \mbox{$\sim 10$\,ms} and only narrowband features, of order $1/\tau$ wide or narrower, 
are expected to be present in the time-shifted estimates of $Y(f)$. 
Note that such narrowband
features could be instrumental/environmental in nature, but they could 
also be genuine GW signals. Since narrowband GW signals are not a
target of this search, our approach is to remove all narrow-band
features from our analysis, regardless of their origin. Further note that
a narrowband feature could also be a statistical fluctuation. In this case,
it would not appear in the analysis with different time-shifts, 
and it would then have
to be included in the search for the SGWB. While the time-shift 
technique is not susceptible to the ``incomplete coverage'' 
issues that are relevant for the IFO-PEM coherence technique,
it cannot identify broadband instrumental correlations. 
Hence, the two techniques are largely complementary.

In order to avoid correlations between estimates with 
different time-shifts $\tau$, the analysis must be performed 
using segment duration $T\le\tau$. Since we are interested in removing \mbox{$\sim 1$\,Hz}
features and narrower, we use time-shifts of order \mbox{1\,s}. This also leads to the choice
\mbox{$T=1$\,s}. Such short segment duration leads to other differences
from the standard SGWB search procedure (where \mbox{$T\sim 100$\,s}), described in
the references~\cite{S4paper}\cite{S3paper}\cite{S1paper}.
In particular, the PSD estimates are obtained using seven neighboring \mbox{1\,s} 
segments on each side (instead of only one), and 
the data quality cuts are imposed
on the PSDs rather than on estimates of $\sigma_Y$, as they are
less susceptible to downward fluctuations in the noise.
We also remove the times identified by the LSC Glitch Group to contain
instrumentally caused transients in the data.
The method was tested using the S5 data between November 2005 and
April 2006, using the calibrated strain data, a \mbox{40-500\,Hz} frequency range, a 
sixth-order butterworth high-pass filter at \mbox{32\,Hz} to remove low-frequency 
detector noise, and 50\% overlapping Hann windows. 
In the study, we never estimated $Y(f)$ without an unphysical time-shift,
to avoid potentially biasing the resulting data-quality cuts.

We define the frequency-dependent signal-to-noise ratio, as a function
of time-shift $\tau$, properly accounting for the frequency bin width $\rmd f$:
\begin{equation} \label{eq:snr}
{\rm SNR_{TS}}(f; \tau) = \frac{|Y(f;\tau)| \; \rmd f}{\sigma(f;\tau) 
\; \sqrt{{\rm d}f}} .
\end{equation}

Using \mbox{1\,s} segments, we performed \mbox{$\pm 1,\pm 2$\,s} time-shifts. We then
attempted three ways of combining these four measurements to estimate the
the SNR at zero-shift, ${\rm SNR_{TS}}(f;0)$: 
Gaussian fit to the four measurements, 
maximum over the four measurements, and average of \mbox{$\pm 1$\,s} measurements.
These three approaches led to very similar results, so in the following we
conservatively use the maximum over the three.

\section{Comparison of the two techniques} \label{sec:relating_techniques}

The PEM-IFO coherence method estimates the environmental contribution
to the coherence between the two interferometers $\gamma_\mathrm{PEM}(f)$.  Coherence $\gamma(f)$ is defined by:

\begin{equation} \label{eq:coherence}
\gamma(f) = \frac{P_{12}(f)}{\sqrt{P_1(f) P_2(f)}}
= \frac{P_{12}(f) Q(f) \rmd f}{\sqrt{P_1(f) P_2(f) Q^2(f) \rmd f / \mathcal{T}} } \; 
\frac{1}{\sqrt{\mathcal{T} \rmd f}} = \frac{{\rm SNR}(f)}{\sqrt{\mathcal{T} \rmd f}} ,
\end{equation}

where $P_{12}(f) = \langle s_1^*(f) s_2(f) \rangle$ is the cross-power spectral density and $\mathcal{T}$ is the total observation time, so $\mathcal{T} \rmd f$ is the total number of segments.  Hence, we
can estimate the environmental contribution to the SNR of the stochastic
search:
\begin{equation}
{\rm SNR_{PEM}}(f) = \gamma_{{\rm PEM}}(f) \sqrt{\mathcal{T} \rmd f} .
\end{equation}

This allows us to directly compare the two techniques, keeping in mind that
${\rm SNR_{TS}}$ is defined using $|Y(f)|$. Moreover, we can define
the environmental contribution to the point estimate, where again,
the PEM subscript indicates that the quantity is estimated using
the PEM-IFO coherence method:
\begin{equation} \label{eq:omega_coherence}
\Omega_{\rm PEM}(f) = \frac{Y_{\rm PEM}(f)}{T} = \frac{\gamma_{\rm PEM}(f)
\sigma(f)}{\sqrt{T}} .
\end{equation}
%

\section{Proposed search algorithm} \label{sec:search_algorithm}

\subsection{Veto} \label{sec:veto}

\begin{figure}
\centering
\includegraphics[width=3.5in]{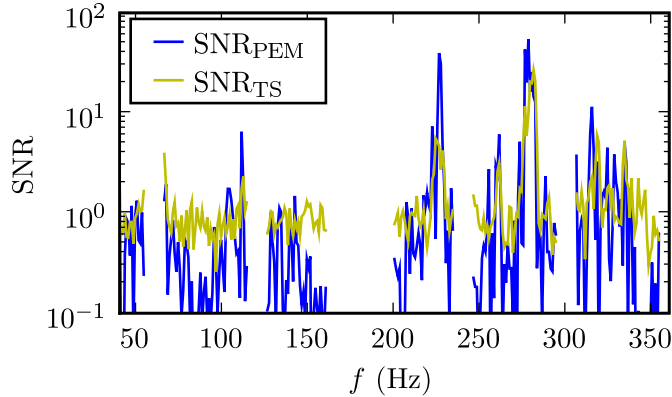}
\caption{Comparison of the PEM-IFO coherence and time-shift techniques for
estimating the contaminated frequency bands. The \mbox{60\,Hz} harmonics 
have already been vetoed.}
\label{fig:comparison}
\end{figure}

The first obvious application of the described techniques is to veto 
frequencies in which H1 and H2 exhibit strong non-gravitational coupling.  
Figure~\ref{fig:comparison} compares the SNR estimates using the two 
techniques. Note that the two techniques generally agree well in identifying
contaminated frequency bands. These bands will be vetoed in the
SGWB search. It is not completely self-evident precisely how this veto 
should be defined, since the IFO channels are coupled to some extent at 
all frequencies. One approach is to veto frequency bins in which both methods' SNRs exceed 2. This 
choice is somewhat relaxed, and it relies on the assumption that 
we can estimate $\Omega_{\rm PEM}$ in the remaining 
frequency bands that pass the veto, as discussed below.
\subsection{Estimate the residual environmental contribution} \label{sec:subtract}

Using equation \ref{eq:omega_coherence}, we can estimate the 
environmental contribution to the estimate of the SGWB amplitude 
$\Omega_{\rm PEM}$, in
the frequency bands that pass the veto described in section~\ref{sec:veto}.
This estimate could be potentially subtracted from the overall 
SGWB amplitude estimate, thereby producing a 
better estimate of the gravitational-wave contribution.

In this case, it is important to understand the uncertainty in the estimate
of $\Omega_{\rm PEM}$. 
The statistical uncertainty in $\Omega_{\rm PEM}$ comes from the
statistical uncertainties on $\gamma_{\rm PEM}$ and on $\sigma_Y$, both of 
which are expected to be negligible compared to the theoretical uncertainty
on the overall amplitude estimate $\sigma_{\Omega} = \sigma_Y/T$. 

\subsection{Estimate systematics} \label{sec:systematics}

The systematic error on $\Omega_{\rm PEM}$
could potentially be large. 
A few major sources of uncertainty are apparent:
the approximation made in the definition of $\gamma_{\rm PEM}$ from
the individual IFO-PEM coherences~\cite{IFOPEMcoherence} and differences in the details of the
data manipulation between the IFO-PEM coherence and SGWB search calculations.
One approach for estimating this systematic error is to compare
the $\Omega_{\rm PEM}$ estimate with the SGWB estimate
in a frequency band that is known to be dominated by environmental
correlations. As the environment dominates, $\Omega_{\rm PEM}$ should explain 
the entire SGWB estimate of $\Omega_0$ in this band. 
The mismatch would give us an estimate on 
the systematic error of our IFO-PEM coherence measurement.

We have performed such a calculation. Our preliminary results indicate
that the IFO-PEM and the SGWB estimates agree within 50\% in the 
contaminated frequency bands. However, we believe we can significantly improve
this agreement; the differences in the data manipulation 
in the two techniques (segment duration, frequency resolution, PSD weighting, etc) were
significant and we expect to minimize them in future iterations.

Another indication of the systematic error could be obtained from hardware 
injections. At semi-regular intervals in the LIGO science runs, we 
injected simulated SGWB signals 
by physically shaking the optics; the SGWB search on these segments tests
how accurately we can recover the injected signal's amplitude.  
If we instead treat the injection data stream as a PEM channel, we
can verify that we can subtract out the injection and instead produce an 
estimate consistent with noise.  
A signal detection would imply an imperfect subtraction.

\subsection{SGWB search} \label{sec:run}

Having identified the most egregiously correlated frequency bands, and 
estimated the environmental contribution $\Omega_{\rm PEM}$, we can
proceed with the usual SGWB search algorithm. This includes a number of 
diagnostics on the quality of data, such as tests of Gaussianity, long-term
stationarity, etc. Potential problems would become apparent here, such
as omissions in the frequency and time vetoes.
Any problems would be investigated and used as input in another iteration of 
the veto and environment contribution estimation steps. The very last step
is to produce the SGWB estimate, and potentially subtract the 
$\Omega_{\rm PEM}$ estimate.

\section{Conclusions} \label{sec:conclusions}

We have presented the time-slide technique, by which all narrowband, correlated
noise between two interferometers can be identified.  It provides 
complementary coverage of
the space of possible non-gravitational couplings between the 
interferometers, alongside the IFO-PEM
coherence technique.  It is important to note that these techniques leave
some blind spots: they leave out the possibility of broad-band 
environmental contributions that are unmonitored in the observatory's
environment, and they misestimate their contributions if the couplings are
non-linear. This includes effects such as seismic upconversion, a
non-linear process by which low-frequency seismic activity excites 
higher-frequency vibrations in the
instrument, and stray light reflections in the beam tubes
that introduce cross-talk between the interferometers.

\ack

We would like to acknowledge many useful discussions
with members of the LIGO Scientific Collaboration stochastic
analysis group which were critical in the formulation of the
methods and results described in this paper. This work has
been supported in part by NSF grants PHY-0200852 and
PHY-0701817. LIGO was constructed by the California Institute of
Technology and Massachusetts Institute of Technology
with funding from the National Science Foundation and operates
under cooperative agreement PHY-0107417. This paper
has the LIGO Document Number LIGO-P070128-00-Z.

\section*{References}
\bibliography{biblio}

\end{document}